\documentclass{article}
\usepackage[preprint]{spconf}
\usepackage{amsmath,graphicx}
\copyrightnotice{\copyright\ IEEE 2023}
\toappear{To appear in {\it Proc.\ ICASSP2023,
                    June 04-10, 2023, Rhodes Island, Greece}}
\usepackage{multirow,multicol}
\usepackage{bbding}
\usepackage{booktabs}
\usepackage{graphicx}
\usepackage{makecell}
\usepackage{mathptmx}
\usepackage{threeparttable}

\title{Adaptable End-to-End ASR Models using Replaceable Internal LMs and Residual Softmax}
\name{Keqi Deng, Philip C. Woodland\thanks{Keqi Deng is funded by the Cambridge Trust. This work has been performed using resources provided by the Cambridge Tier-2 system operated by the University of Cambridge Research Computing Service (www.hpc.cam.ac.uk) funded by EPSRC Tier-2 capital grant EP/T022159/1.}}
\address{Department of Engineering, University of Cambridge, Trumpington St., Cambridge, UK.\\\small{\texttt{\{kd502, pcw\}@eng.cam.ac.uk}}}

\begin{document}
%
\maketitle
\begin{abstract}
End-to-end (E2E) automatic speech recognition (ASR) 
implicitly learns the token sequence distribution of paired audio-transcript training data.
However, it still suffers from domain shifts from training to testing, and domain adaptation is still challenging. To alleviate this problem, this paper designs a replaceable internal language model (RILM) method, which makes it feasible to 
directly replace the internal language model (LM) of E2E ASR models with a target-domain LM in the decoding stage when a domain shift is encountered.
Furthermore, this paper proposes a residual softmax (R-softmax) that is designed for CTC-based E2E ASR models to adapt to the target domain without re-training during inference.
For E2E ASR models trained on the LibriSpeech corpus, experiments showed that the proposed methods gave a 2.6\% absolute WER reduction on the Switchboard data and a 1.0\% WER reduction on the AESRC2020 corpus while maintaining intra-domain ASR results.

\end{abstract}
\vspace{-0.25cm}
\section{Introduction}
\label{sec:intro}
\vspace{-0.15cm}
End-to-end (E2E) automatic speech recognition (ASR) models simplify conventional pipeline ASR methods and directly transcribe input speech into corresponding text \cite{8068205,6638947}. Due to a large amount of labelled training data, E2E ASR models surpass pipeline methods on most public datasets \cite{9688009}.
However, E2E ASR still suffers from unseen domains \cite{tsunoo22_interspeech}, and large quantities of labelled data are not always feasible to collect and can therefore be limited \cite{9746480}. Adaptation training methods can be utilised to alleviate this issue when the target domain has enough paired data \cite{tsunoo19_interspeech, 6424251}. However, text-only data from the target domain is easier to obtain in most scenarios, and it is more efficient to bias the E2E ASR systems to the target domain using such data \cite{tsunoo22_interspeech}.

There are several studies exploring the use of text-only data via an external language model (LM). Shallow fusion \cite{chorowski2015attention} which linearly interpolates the E2E ASR with an external LM is straightforward and widely deployed \cite{8462682}.
Several structural fusion methods like deep fusion \cite{gulcehre2015using} and cold fusion \cite{sriram18_interspeech} have been proposed, but require additional training and haven't replaced shallow fusion \cite{chorowski2015attention} as the dominant LM integration method \cite{9003790, 9383515}.
Considering that E2E models learn to  model the context between words and characterise the training data of the source domain, a density ratio method \cite{9003790} was proposed as an extension of shallow fusion. It subtracts the score of a source-domain LM from the log-linear combination of the E2E ASR model and target-domain LM scores \cite{9003790}. Furthermore, the estimate of the E2E ASR model's internal LM\footnote{We refer to the ability of E2E ASR to model the context of a token sequence as an internal LM \cite{9383515}.} has been explored \cite{9383515, 9415039, 9746948, zeineldeen21_interspeech}.
However, both the density ratio and internal LM estimation methods
make the decoding process more complicated, and it is not always feasible to accurately estimate the internal LM due to domain mismatch \cite{tsunoo22_interspeech}. To simplify the decoding process based on the internal LM estimation,
\cite{tsunoo22_interspeech} further proposed a residual LM that
models the residual factor of external and internal LMs, but the internal LM estimate must be pre-calculated for all text data before training and the issue of inaccurate estimation still exists.
Work in \cite{meng22_interspeech} explores fine-tuning the internal LM with text-only data but requires regularisation strategies to avoid the internal LM over-learning target domains.

Previous methods rely on an external LM with the estimate of the internal LM used to bias the prediction of E2E ASR systems, thus improving the cross-domain ASR performance. However, incorporating the external LM demands extra computational cost and additional parameters, and accurate internal LM estimation is not always feasible \cite{tsunoo22_interspeech}. In this paper, the motivation is making the E2E ASR system itself adaptable and can be biased to unseen domains without needing an external LM or re-training. Therefore,
this paper proposes a replaceable internal LM (RILM) method, through which the internal LM of the E2E ASR system can be directly replaced with a target-domain LM during the decoding stage to improve the cross-domain ASR accuracy. Furthermore, this paper designs a residual softmax (R-softmax) for CTC-based ASR models that efficiently adapts to the target domain during the inference stage. With E2E ASR models 
trained on the LibriSpeech corpus \cite{7178964}, experiments showed that the proposed methods greatly boosted the cross-domain ASR accuracy on the Switchboard \cite{225858} and AESRC2020 \cite{9413386} corpus while performing robustly in intra-domain scenarios.

\begin{figure*}[t]
    \centering
    \includegraphics[width=0.92\linewidth]{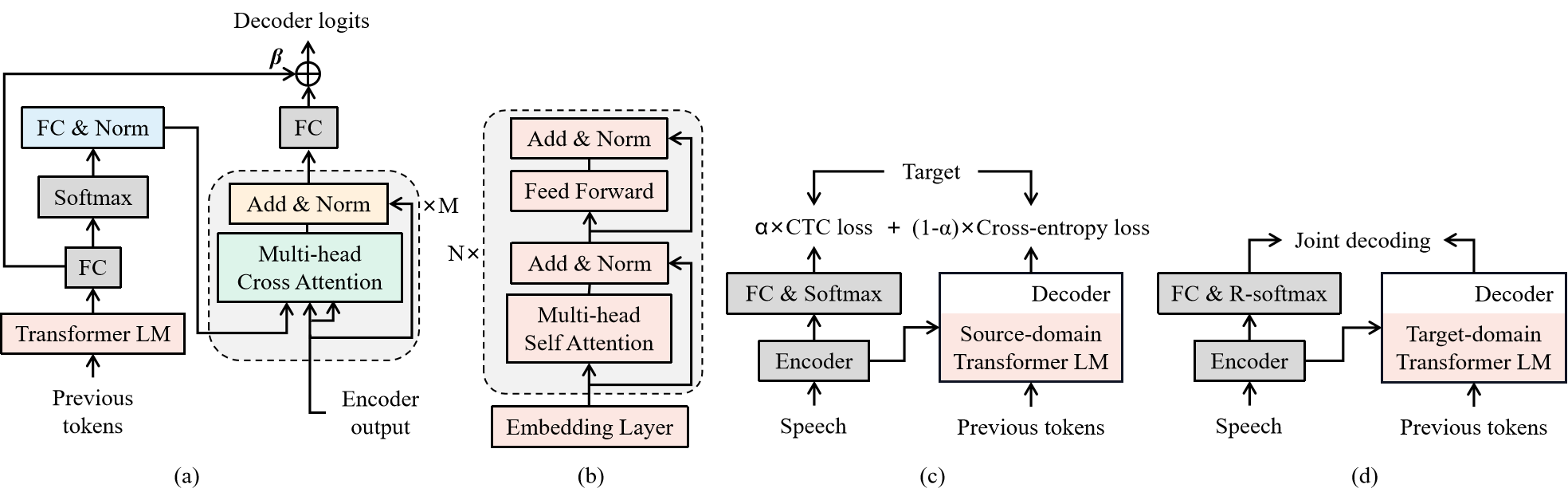}
    \caption{Illustration of the proposed methods. (a) a decoder based on proposed RILM method; (b) structure of Transformer LM that is part of the decoder in Fig.1 (a); (c) the training and (d) the decoding process of the proposed E2E ASR system.}
    \label{fig:arch}
\end{figure*}
The rest of this paper is organized as follows:
Sec.~\ref{sec:format} describes the proposed methods.
Sec.~\ref{sec:typestyle}
details the data, model, and experimental results. Conclusions are provided in Sec.~\ref{sec:print}.
\section{Proposed Methods}
\label{sec:format}
This paper proposes a replaceable internal language model (RILM) method to make the attention-based encoder-decoder E2E ASR system itself adaptable when encountering domain shift. In addition, this paper further designs a residual softmax (R-softmax) for the CTC-based models in cross-domain scenarios. 
In this paper, a hybrid CTC/attention framework \cite{8068205} is employed to utilise both proposed methods.
The realisation of the methods is shown in Fig.~\ref{fig:arch}, where FC represents a fully connected layer, and \textcircled{+} denotes addition operations.
\subsection{Replaceable Internal Language Model (RILM)}
A standard Transformer decoder contains a stack of several identical layers \cite{Vaswani2017}, and each layer consists of several sub-layers, including a self-attention module, a cross-attention module, and a feed-forward module. It is the self-attention module in each layer that enables the decoder to model the context between token sequences and is therefore interpreted as having an internal LM. However, the cross-attention module in each layer makes the decoder dependent on the acoustic encoder output and thus can not be separately pre-trained on text data \cite{9688009}, and also makes the internal LM hard to replace.

Inspired by prior work \cite{9688009}, the proposed RILM method makes some modifications to the Transformer decoder structure. As shown in Fig.~\ref{fig:arch} (a) and (b), the RILM method only retains the cross-attention module in the last $M$ layers and removes it from the previous $N$ layers.
The structure of the previous $N$ layers without cross-attention mechanisms is the same as a standard Transformer LM, and is named the internal LM of the E2E ASR system decoder in this paper, as it is the self-attention module that makes it feasible for the decoder to model the context between tokens.
To avoid a mismatch when directly replacing the internal LM, the RILM method first calculates a predicted distribution of the internal LM and then uses a fully connected layer to transform its dimensions to be the same as the attention module before feeding it into the following cross-attention modules.

Inspired by \cite{he2016deep}, a highway connection can be used to directly connect the internal LM to the final output, thus emphasizing semantic information. The final output of the model is as follows and is illustrated in Fig.1 (a):
\begin{equation}
{\rm logits} = {\rm logits}_{A}+\beta \cdot{\rm logits}_{L} \label{logits}
\end{equation}
where ${\rm logits_L}$ and ${\rm logits_A}$ denote the output of the internal LM and the cross-attention modules, respectively. $\beta$ is a tunable hyperparameter and is shown in Fig.~\ref{fig:arch} (a).

\subsection{Residual Softmax (R-softmax)}
Previous work \cite{tsunoo22_interspeech, 9383515, 9003790, 9415039} focused on the encoder-decoder E2E ASR structure rather than CTC-based structure to estimate the internal LM because CTC-based models are generally not considered capable of modelling context between output tokens due to conditional independence assumption.

However, CTC-based E2E ASR models learn the training data distribution and are affected by the frequency of words in the training data \cite{DBLP:journals/taslp/DengCYY22}. The CTC-based model therefore at least has the modelling ability of a unigram LM, and this paper aims to adapt it to the target domain effectively without re-training during inference. Therefore, this paper proposes the R-softmax for CTC-based models. 

Let $x$ and $y$ be input data and its label, respectively, and $p^{t}_j$ and $p^{s}_j$ be the frequency of the $j$-th token in target-domain and source-domain data distribution, respectively.
Assume $\phi$ is a desired predicted CTC probability of the target domain, with form $\phi_{j}=p(y=j|x)=(p(x|y=j)/{p(x)})\cdot p(y=j)=(p(x|y=j)/{p(x)})\cdot p^{t}_j$, 
and $\hat{\phi}$ is a predicted CTC probability of the source domain, 
with form $\hat{\phi_{j}}=\hat{p}(y=j|x)=(p(x|y=j)/{\hat{p}(x)})\cdot\hat{p}(y=j)=(p(x|y=j)/{\hat{p}(x)})\cdot p^{s}_j$ \footnote{We assume that all data in the source and target domains are generated from the same process $p(x|y=j)$ following \cite{ren2020balanced}.}. Let $l_j$ be a logit of the $j$-th token output by a CTC-based model, then $\hat{\phi_{j}}={\rm exp}({l_j})/{\sum_{i=1}^{V}{\rm exp}(l_i)}$, where V denotes the vocabulary size. If $\hat{\phi_{j}}$ is expressed by the normal softmax function, then ${\phi_{j}}$ can be obtained via the R-softmax as follows\footnote{The key to the proof for R-softmax is to first show that $\sum_{i=1}^{V}{\rm exp}(l_i)=\sum_{i=1}^{V}{\rm exp}(l_i-{\rm log}({\hat{\phi_i}}/{\phi_i}))$ and then substitute it into the denominator of $\phi_j = {{\rm exp}(l_j - {\rm log}(\hat{\phi_j}/\phi_j))}/{\sum_{i=1}^{V}{\rm exp}(l_i)}$.}:
\begin{equation}
\phi_{j} = \frac{{\rm exp}(l_j)\cdot(p_j^t/p_j^s)}{\sum_{i=1}^{V}{\rm exp}(l_i)\cdot(p_i^t/p_i^s)} \label{r-softmax}
\end{equation}

Considering that some tokens may never appear in the corresponding text, the R-softmax employs a smoothing strategy when counting the $p^{t}_j$ and the $p^{s}_j$ as follows:
\begin{equation}
p_i=
\begin{Large}
\begin{cases}
\frac{ C_i}{C}-\frac{{\rm I}_{\{n_0 \neq 0\} }}{(V-n_0)\times C},& \text{\normalsize $ C_i >0$ } \\
\frac{{\rm I}_{\{n_0\neq 0\}}}{n_0 \times C},& \text{\normalsize otherwise} \label{tlce2}
\end{cases}
\end{Large}
\end{equation}
where $C$ denotes the total number of counts for all tokens, $C_i$ is the count of the $i$-th token, $n_0$ represents the number of tokens that never appear, and ${\rm I}_{\{n_0 \neq 0\}}$ is an indicator function which is 1 if $n_0\neq 0$ and 0 otherwise.

Furthermore, the ${\rm [blank]}$ label in CTC is unique and never appears when counting the frequency of training tokens.
Therefore, the R-softmax keeps the predicted probability of ${\rm [blank]}$ the same as when using the normal softmax.
Assume the index of ${\rm [blank]}$ as 1 and the weight ($p_1^t/p_1^s$) of ${\rm [blank]}$ in R-softmax as $k$, the value of $k$ can be calculated using the following equations:
\begin{equation}
    \frac{k \cdot {\rm exp}({l}_{1})}{k \cdot {\rm exp}({l}_{1})+\sum_{i=2}^{V}(p_i^t/p_i^s)\cdot {\rm exp}({l}_{i})}=\frac{{\rm exp}({l}_{1})}{\sum_{i=1}^{V}{\rm exp}({l}_{i})}
\end{equation}
\begin{equation}
    \frac{p_1^t}{p_1^s}=k=\frac{\sum_{i=2}^{V}(p_i^t/p_i^s)\cdot {\rm exp}({l}_{i})}{\sum_{i=2}^{V}{\rm exp}({l}_{i})} \label{ctc-soft}
\end{equation}
\section{Experiments}
\label{sec:typestyle}
\subsection{Corpus}
\label{ssec:corp}
E2E ASR models were trained on Librispeech corpus \cite{7178964}, a read speech corpus that contains 960-hour labelled speech data, and the standard test sets from Librispeech (“test-clean/-other”) were used for intra-domain evaluation.
To verify the effectiveness of the proposed methods on domain adaptation, 
two out-of-domain datasets were used in the experiments. The first was the dev and test sets (Eval2000 and RT03) normally used with the Switchboard (SWBD) telephone conversation \cite{225858} corpus. 
The text data for
target-domain LM training was the SWBD and
Fisher transcriptions.
Also, the dev and test sets from the AESRC2020 corpus \cite{9413386} were used, which is an accented English speech corpus, and the text data for
target-domain LM training was the training transcriptions.

\subsection{Model Descriptions}
Baseline models and the proposed E2E ASR systems were built using the ESPnet2 toolkit \cite{watanabe2018espnet}. Experiments used 80-dimensional filter bank features.
Text output used 5000 modeling units, including 4997 byte pair encoding (BPE) units \cite{gage1994} and 3 non-verbal symbols.

Under a hybrid CTC/attention framework, a Conformer \cite{gulati20_interspeech} baseline model was developed following the ESPnet2 recipe \cite{watanabe2018espnet} (i.e., 12-layer Conformer encoder and 6-layer Transformer decoder with 512 attention dimensions, 2048 feed-forward dimensions, and 8 heads). To achieve a more competitive performance by including popular self-supervised pre-training, we also built a strong Wav2vec2.0-based \cite{NEURIPS2020_92d1e1eb, hsu21_interspeech} baseline model to replace the Conformer encoder with a Wav2vec2.0 encoder \cite{hsu21_interspeech} provided by Fairseq \cite{ott2019fairseq} (i.e. "w2v\_large\_lv\_fsh\_swbd\_cv"). It also used a fully connected layer to change the dimensions of the encoder output from 1024 to 512 before it was fed into the decoder, and the decoder was the same as the previous baseline model.
For the proposed E2E ASR systems shown in Fig.~\ref{fig:arch}, the decoder contained 6 self-attention, cross-attention, and feed-forward modules (i.e., $M$=$N$=$6$), the FCs respectively changed the dimensions to 5000 and 512, and other structures were the same as the baseline models. The $\beta$ in Eq.~\ref{logits} was set to 0.3 for Conformer-based model and 0 for Wav2vec2.0-based models.


During training, the CTC weight was 0.3, and the models based on Conformer and Wav2vec2.0 were trained for 35 and 20 epochs respectively. A source-domain internal Transformer LM (6 layers) was trained on Librispeech LM corpus \cite{7178964} for 25 epochs following the ESPnet2 recipe, and it was fine-tuned on target-domain text corpus for extra 5 epochs. The source-domain LM was used as the internal LM part of the proposed decoder and was fixed during ASR training.
Model parameters from the last 10 epochs were averaged to avoid over-fitting. 
During decoding, CTC weights were set to 0.3 and 0.5 for models based on Conformer and Wav2vec2.0 encoders respectively.
We implemented shallow fusion \cite{chorowski2015attention} with 0.1 external LM weight and density ratio \cite{9003790} with 0.2 target and 0.1 source LM weights following \cite{9383515} to compare with the proposed methods. The beam size was 20. 

\subsection{Experimental results}
Experiments were conducted to compare the proposed E2E ASR methods with the strong baseline models for both intra-domain and cross-domain scenarios. 
This paper aims to improve cross-domain accuracy while avoiding a degradation of intra-domain performance.
\label{sec:majhead}
\begin{table}[t]
  \caption{WER of the E2E ASR models in intra-domain scenarios on the test sets of Librispeech.}
  \label{tab:librispeech}
  \centering
  \setlength{\tabcolsep}{2.5mm}
  \begin{tabular}{l c| c c}
    \Xhline{3\arrayrulewidth}
    {Model}&{Encoder} &{Test-clean}&{Test-other} \\
    \hline
    Baseline&Conformer &2.7& 6.8\\
    Proposed& Conformer &\textbf{2.7}&\textbf{6.5}\\
    \hline
    Baseline&Wav2vec2.0 &3.6&5.5 \\
    Proposed& Wav2vec2.0&\textbf{2.2}&\textbf{4.6}\\
    \Xhline{3\arrayrulewidth}
  \end{tabular}
\end{table}
\begin{table}[t]
\vspace{-0.3cm}
  \caption{WER of the E2E ASR models in cross-domain adaptation scenarios on the dev and test sets of AESRC2020.}
  \label{tab:aesrc}
  \centering
  \setlength{\tabcolsep}{1.0mm}
  \begin{tabular}{l c c c| c c}
    \Xhline{3\arrayrulewidth}
    \multirow{2}{*}{Model}&\multirow{2}{*}{Encoder}&Replace&Residual &\multirow{2}{*}{Dev}&\multirow{2}{*}{Test} \\
     & &Internal LM&Softmax && \\
    \hline
    Baseline&Conformer &\XSolidBrush& \XSolidBrush &14.3& 15.8\\
    Proposed& Conformer&\Checkmark&\Checkmark &\textbf{13.4}&\textbf{14.8}\\
    Proposed& Conformer&\XSolidBrush&\Checkmark &13.7& 15.2\\
    Proposed& Conformer&\Checkmark&\XSolidBrush &13.8& 15.2\\
    Proposed& Conformer&\XSolidBrush&\XSolidBrush &14.1& 15.6\\
    \hline
    Baseline&Wav2vec2.0 &\XSolidBrush& \XSolidBrush &11.6&12.4 \\
    Proposed& Wav2vec2.0&\Checkmark&\Checkmark &\textbf{10.7}&\textbf{11.4}\\
    Proposed& Wav2vec2.0&\XSolidBrush&\Checkmark &11.0& 11.8\\
    Proposed& Wav2vec2.0&\Checkmark&\XSolidBrush &11.2&12.0 \\
    Proposed& Wav2vec2.0&\XSolidBrush&\XSolidBrush &11.3&12.2\\
    \Xhline{3\arrayrulewidth}
  \end{tabular}
  \vspace{-0.3cm}
\end{table}
\begin{table}[t]
  \caption{WER of the E2E ASR models in cross-domain adaptation scenarios on the dev and test sets of SWBD. Dev set was obtained following ESPnet2 \cite{watanabe2018espnet} processing.}
  \label{tab:swbd}
  \centering
  \setlength{\tabcolsep}{0.3mm}
  \begin{tabular}{l c c c| c c c}
    \Xhline{3\arrayrulewidth}
    {\multirow{2}{*}{Model}}&
    \multirow{2}{*}{Encoder}&Replace&Residual &\multirow{2}{*}{Dev}&\multirow{2}{*}{Eval2000}&\multirow{2}{*}{RT03} \\
     & &ILM&Softmax &&& \\
    \hline
    Baseline&Conformer &\XSolidBrush& \XSolidBrush &39.7& 33.6& 38.2\\
    Proposed& Conformer&\Checkmark&\Checkmark &\textbf{37.1}&\textbf{31.2}& \textbf{35.9}\\
    Proposed&Conformer&\XSolidBrush&\Checkmark &38.0& 31.6& 36.4\\
    Proposed&Conformer&\Checkmark&\XSolidBrush &37.9& 32.0& 36.7\\
    Proposed& Conformer&\XSolidBrush&\XSolidBrush &38.7& 32.3& 37.2\\
    \hline
    Baseline&Wav2vec2.0 &\XSolidBrush& \XSolidBrush &27.8&18.3 &21.1 \\
    Proposed& Wav2vec2.0&\Checkmark&\Checkmark &\textbf{26.1}&\textbf{17.8}& \textbf{20.4}\\
    Proposed& Wav2vec2.0&\XSolidBrush&\Checkmark &26.6& 18.1& 20.9\\
    Proposed&Wav2vec2.0&\Checkmark&\XSolidBrush &27.3& 18.3&21.3 \\
    Proposed& Wav2vec2.0&\XSolidBrush&\XSolidBrush &27.7& 18.5& 21.6\\
    \Xhline{3\arrayrulewidth}
  \end{tabular}
\end{table}
\begin{table}[t]
\vspace{-0.25cm}
  \caption{WER for proposed methods and shallow fusion (SF) and density ratio (DR) in cross-domain scenarios on AESRC and SWBD. Wav2vec2.0 encoder was used.}
  \vspace{0.05cm}
  \label{tab:other}
  \centering
  \setlength{\tabcolsep}{1.0mm}
  \begin{tabular}{l c c| c c c}
    \Xhline{3\arrayrulewidth}
    \multirow{2}{*}{Model}&\multicolumn{2}{c|}{AESRC2020} &\multicolumn{3}{c}{SWBD}\\
    &Dev&Test&Dev&Eval2000&RT03\\
    \hline
    Baseline + SF \cite{chorowski2015attention}$^*$&10.9&11.8&27.4&17.9&20.5\\
    Baseline + DR \cite{9003790}$^*$&10.8&11.6&27.3&\textbf{17.8}&\textbf{20.4}\\
    Proposed &\textbf{10.7}&\textbf{11.4}&\textbf{26.1}&\textbf{17.8}&\textbf{20.4}\\
    \hline
    \Xhline{3\arrayrulewidth}
  \end{tabular}
  \begin{tablenotes}
  \footnotesize
  \item{\hspace{-3.5mm}*}{Note that SF and DR rely on external LMs, which are trained on \\ \hspace{-3.5mm} source/target-domain text.}
  \end{tablenotes}
  \vspace{-0.3cm}
\end{table}
\subsubsection{Intra-domain ASR}
The results in Table~\ref{tab:librispeech} showed that the proposed ASR system could slightly outperform the baseline model in the source domain when using the Conformer encoder. In addition, due to teacher forcing \cite{Williams1989} in decoder training,
some overfitting was observed on the baseline model when using the stronger Wav2vec2.0 encoder, while the proposed ASR system performs more robustly, which may be due to the fixed internal LM of RILM during ASR training.

\subsubsection{Cross-domain ASR}
Experiments were then conducted to compare the cross-domain ASR performance on the AESRC2020 and SWBD corpora.
As shown in Table~\ref{tab:aesrc} on the AESRC2020 corpus, without needing an external LM or re-training, the proposed methods outperformed the baseline models by about 1\% absolute WER reduction in the cross-domain scenarios whether using the Conformer or Wav2vec2.0 encoders. More specifically, in Table~\ref{tab:aesrc}, ablation studies showed that both the proposed RILM method and R-softmax could bring around 0.5\% absolute reduction in WER and
were effective for flexible domain adaptation. It also shows that the proposed methods has broad application, whether for CTC-based, encoder-decoder, or hybrid CTC/attention models.

The results on the SWBD corpus in Table~\ref{tab:swbd} showed more severe performance degradation, which was also due to different corpus collection environments.
However, after using the robust Wav2vec2.0 encoder \cite{hsu21_interspeech}, the performance degradation was greatly reduced. The conclusion from SWBD was in general consistent with that of the AESRC2020 corpus: 1. the proposed methods achieved better cross-domain performance compared with the baseline models, with around 2.6\% absolute WER reduction for Conformer encoder and about 1.7\% for Wav2vec2.0 encoder; 2. both the designed RILM and R-softmax worked for ASR models in cross-domain scenarios and had broad application.

We also compared the proposed methods with shallow fusion \cite{chorowski2015attention} and density ratio \cite{9003790} methods on the AESRC2020 and SWBD corpora, and the results in Table~\ref{tab:other} showed that the proposed methods achieved close cross-domain
results to them, but is more flexible as it does not use an external LM.
\section{Conclusions}
\label{sec:print}
This paper proposes a replaceable internal LM (RILM) method, which effectively makes the E2E ASR system itself adaptable and can be biased to target domains without needing an external LM or re-training. Through the proposed RILM method, the E2E ASR model decoder internal LM can be directly replaced by a target-domain LM, thus achieving flexible adaptation. Furthermore, this paper designs a residual softmax (R-softmax) for CTC-based models to achieve domain adaptation without re-training during inference. Under a hybrid CTC/attention framework, both proposed methods are employed in this paper. Experimental results showed that the proposed methods greatly improved cross-domain ASR performance while maintaining intra-domain results.

\small
\bibliographystyle{IEEEbib}
\bibliography{strings,refs}

\end{document}